\documentclass[a4paper,11pt]{article}
\pdfoutput=1 % if your are submitting a pdflatex (i.e. if you have
\usepackage{jcappub} % for details on the use of the package, please
\usepackage[T1]{fontenc} % if needed
\usepackage[sort&compress]{natbib}
\newcommand{\be}{\begin{equation}}
\newcommand{\ee}{\end{equation}}
\newcommand{\nn}{\nonumber}
\newcommand{\ba}{\begin{eqnarray}}
\newcommand{\ea}{\end{eqnarray}}
\renewcommand\[{\begin{equation}}
\renewcommand\]{\end{equation}}
\newcommand{\Mc}{{\cal M}}

\title{\boldmath Defocusing of Null Rays in Infinite Derivative Gravity}

\author[a,1]{Aindri\'u Conroy,}
\author[b,c,2]{Alexey S. Koshelev}
\author[a,3]{and Anupam Mazumdar}

\affiliation[a]{Consortium for Fundamental Physics, Lancaster University, Lancaster, LA1 4YB, UK}
\affiliation[b]{Departamento  de F\'isica and Centro  de  Matem\'atica  e 
Aplica\c c\~oes,  Universidade  da  Beira  Interior,  6200  Covilh\~a, 
Portugal}
\affiliation[c]{Theoretische Natuurkunde, Vrije Universiteit Brussel and The 
International Solvay Institutes, Pleinlaan 2, B-1050, Brussels, Belgium}

\emailAdd{a.conroy@lancaster.ac.uk}
\emailAdd{alexey@ubi.pt}
\emailAdd{a.mazumdar@lancaster.ac.uk}

\abstract{
Einstein's General theory of relativity permits spacetime singularities, where null geodesic congruences
 \emph{focus} in the presence of matter,  which satisfies an appropriate energy condition. In this paper, we provide a 
 {\it minimal defocusing condition} for null congruences without assuming any {\it ansatz}-dependent background solution.
The two important criteria are: (1) an additional scalar degree of freedom, besides the massless graviton must be  introduced into the spacetime; and (2) an \emph{infinite} derivative theory of gravity  is required in order to avoid tachyons or ghosts in the  graviton propagator. In this regard, our analysis strengthens earlier arguments for constructing non-singular bouncing cosmologies within an infinite derivative theory of gravity, without assuming any {\it ansatz} to solve the full equations of motion.}

\begin{document}
\maketitle
\flushbottom

\section{Introduction}

The classical theory of General relativity (GR) admits spacetime singularities, see~\cite{Hawking:1973uf,Wald:1984rg} and \cite{Penrose:1964wq}. Such a singularity manifests itself, at short distances, as a blackhole and, at small time scales, as the cosmological Big Bang singularity. In particular, the latter has a detrimental impact on the construction of an ultraviolet (UV) complete theory of primordial inflation~\cite{Borde:1993xh}. 

A major shortcoming of GR is that it may never evade the {\it shadow} of the  cosmological singularity problem, unless one violates a relevant energy condition. The appropriate energy condition depends on the nature of the geodesic congruences involved. For null rays, one requires a violation of the \emph{Null Energy Condition} (NEC), i.e. $T_{\mu\nu}k^{\mu}k^{\nu}\geq 0$, where $\mu,~\nu$ run from $0,~1,~2,~3$ in 4-dimensions, $k^{\mu}$ is a null tangent vector, and $T_{\mu\nu}$ is the energy momentum tensor.  
For timelike geodesics, one has to violate the \emph{Weak Energy Conditions} (WEC), i.e. $T_{\mu\nu}\xi^{\mu}\xi^{\nu}\geq 0$, where $\xi^{\mu}$ is 
timelike. The famous Hawking-Penrose Singularity Theorems~\cite{Penrose:1964wq} were derived in the context of null and timelike versions of the Raychaudhuri equation~\cite{Raychaudhuri:1953yv}~\footnote{The definition of a singularity which forms the basis of the Hawking-Penrose theorems concerns geodesic completeness and it is this notion of a spacetime containing causal (timelike or null) geodesic congruences which focus, (or indeed \emph{defocus}) that interests us here. If a spacetime is causally incomplete, then a freely falling photon passing along this geodesic will cease to exist in a finite `time' (affine parameter). As such, we can reasonably call this a \emph{singular} spacetime, see \cite{Wald:1984rg}, \cite{Geroch:1968ut}, and \cite{Albareti:2014dxa}.}
%~\footnote{Throughout our discussion, we will deal exclusively with null geodesics congruences as opposed to the timelike case. This is because, in a geometrically flat spacetime, the null convergence condition is more easily satisfied than its timelike counterpart, and as a result is a stronger test in terms of defocusing. See \cite{Albareti:2014dxa} and \cite{Geroch:1968ut} for further details.}.

At time scales close to the Planck scale, i.e. $M_P\sim 2.4\times 10^{18}$~GeV, one would  expect the Einstein-Hilbert action to be modified by higher curvature corrections, i.e. higher derivative modifications made up of the Ricci scalar, Ricci tensor and the Riemann/Weyl terms.  However, higher derivative theories are beset by classical and quantum instabilities. For instance, quadratic curvature gravity, which is renormalizable, succeeds in improving the UV behaviour, but at the expense of introducing 
{\it ghosts} in the spin-2 component of the graviton propagator~\cite{Stelle}. The presence of ghosts is not a welcome sign as it renders the vacuum unstable at both classical 
and at quantum level.

Recently, the issue of ghosts in the spin-2 component of quadratic curvature gravity has been addressed in ~\cite{Biswas:2011ar}, where it was found that one requires an infinite number of covariant derivatives acting on the curvature to maintain general covariance and avoid the addition of ghosts at a perturbative level. Although the graviton propagator is modified by these infinite derivatives, it is still possible to retain the original massless graviton degrees of freedom, if one requires that \emph{no new poles} are introduced in either the spin-0 or spin-2 component of the graviton propagator.
This can be achieved if the propagator is modified by an {\it exponent of an entire function} of the d'Alembertian (denoted as $\Box= g^{\mu\nu}\nabla_{\mu}\nabla_{\nu}$), which contains no roots, by construction~\cite{Biswas:2011ar,Tomboulis,Tseytlin:1995uq,Siegel,Biswas:2005qr,Modesto}.  For a further {\it covariant generalisation}, see~\cite{Edholm:2016hbt}.

Typically, such an exponential function in the propagator weakens the classical and quantum effects of gravity in the UV. For instance, it was found, at the linearised limit, that there are no blackhole-like singularities in a static spacetime~\cite{Biswas:2011ar}, nor a time-dependent collapse of matter~\cite{Frolov:2015bia}. It was also observed that the gravitational entropy~\cite{Wald-1}, 
of a static and axisymmetric metric receives zero contribution from the infinite derivative sector of the action, when no additional scalar propagating modes are introduced. In this case, the gravitational entropy is strictly given by the {\it area-law},
arising solely from the contribution of the Einstein-Hilbert action~\cite{Conroy:2015wfa}. In the quantum context, the interplay between an 
{\it exponentially enhanced} vertex operator and an {\it exponentially suppressed } graviton propagator ensures that, beyond 1-loop, the theory becomes finite~\cite{Tomboulis,Modesto,Talaganis:2014ida}.
Furthermore, it was conjectured that the high energy scattering of gravitons in infinite derivative theories lead to a finite amplitude, which does not become arbitrarily large for given external momenta~\cite{Talaganis:2016ovm}. 
A different and rather intriguing motivation to investigate infinite derivative theories of gravity
arises from  string theory (ST)~\cite{Polchinski:1998rr} and string field theory (SFT) models, which appear in the context of  
noncommutative geometry \& SFT~\cite{Witten:1985cc},  for a review, see \cite{Siegel:1988yz}, and various {\it toy model}s of SFT such as $p$-adic strings~\cite{Freund:1987kt,Freund:1987ck,Brekke:1988dg,Frampton:1988kr}, zeta strings~\cite{Dragovich:2007wb}, and strings quantized on a random lattice~\cite{Douglas:1989ve,Gross:1989ni,Brezin:1990rb,Ghoshal:2006te,Biswas:2004qu}. A key feature of these models is the presence of an {\it  infinite series of higher-derivative} series in $\alpha'$, the inverse of string tension, incorporating non-locality in the form of an {\it exponential kinetic} correction. Similar infinite-derivative modifications have also been argued to arise in the asymptotic safety approach to quantum gravity~\cite{Krasnov:2006du,Krasnov:2007uu}, see for a review~\cite{Ambjorn:2012jv}.

Note that in GR, even in an 
{\it asymptotically-flat Minkowski spacetime}, if we perturb matter and curvature while satisfying the NEC, the null congruences will {\it always} converge, in a finite time. The same holds true for timelike geodesics congruences. However, as null rays more readily converge than their timelike counterparts in a geometrically-flat spacetime, it is preferable for our aims to analyse the defocusing of null geodesic congruences rather than timelike. See \cite{Albareti:2014dxa} and \cite{Geroch:1968ut} for details. The prime question is then: \emph{Can null rays ``defocus'' in an infinite derivative theory of gravity (IDG), which is also ghost-free?}

The aim of this paper is to study the time-dependent singularity problem within the ghost-free, infinite derivative theory of gravity proposed in 
Refs.~\cite{Biswas:2005qr,Biswas:2011ar}, where a non-singular bouncing solution for an infinite derivative equation of motion was obtained through an {\it Ansatz}-led approach. In this respect our current analysis is quite different from earlier investigations in the context of {\it IDG} theories~\cite{Biswas:2005qr}, and \cite{Conroy:2014dja}. Indeed, our main focus  here is to go beyond a specific Ansatz-dependent background solution, and investigate under what
generic conditions an {\it IDG} theory can potentially lead to geodesic completeness, for time-dependent scenarios.

In Section \ref{RC}, we discuss the Raychaudhuri equation in GR, before briefly introducing IDG and its relevant properties, such as dynamical degrees of freedom, see Section \ref{IDG}. In Section \ref{DNC}, we discuss the defocusing of null congruences for ghost-free, infinite derivative theories of gravity, while comparing this behaviour with a well-known finite, quadratic curvature model of inflation. In the final section, we conclude our main results.

\section{Raychaudhuri's equation in General Relativity}\label{RC}

 Let us begin our discussion within GR, where the dynamics of null rays can be understood in a {\it model independent} way by studying 
the Raychaudhuri Equation (RE) for null geodesic congruences, such that $k^{\mu}k_{\mu}=0$, where $k_{\mu}$ is a four vector tangential to the null geodesic congruence, the metric signature is mostly positive, i.e. $(-,+,+,+)$, and $\theta$ is the expansion parameter, which is defined by $\theta= \nabla_{\mu}k^{\mu}$. 

In the simplest case, if we consider the congruence of null rays to be orthogonal to the hypersurface, then the \emph{twist} tensors vanish, and 
we may also neglect the shear tensor, which is
purely spatial, thus making a positive contribution to the r.h.s. of the RE. Thus, we have:
\be\label{eq1}
%\frac{d\theta}{d\tau}+\frac{1}{2}\theta^{2}=-\sigma_{\mu\nu}\sigma^{\mu\nu}+\omega_{\mu\nu}\omega^{\mu\nu}-R_{\mu\nu}k^{\mu}k^{\nu}
 \frac{d\theta}{d\lambda}+\frac{1}{2}\theta^{2}\le-R_{\mu\nu}k^{\mu}k^{\nu},
 \ee
where $\lambda$ is the affine parameter,  and $R_{\mu\nu}$ 
is the Ricci tensor, see~\cite{Wald:1984rg}. In GR, the Einstein equation yields, $$G_{\mu\nu}=\kappa T_{\mu\nu},$$ where $\kappa =8\pi G=M_{P}^{-2}$, which implies $R_{\mu\nu}=\kappa( T_{\mu\nu}-\frac{1}{2} g_{\mu\nu} T)$. 

By contracting with the vector fields $k^{\mu}k^{\nu}$ yields $R_{\mu\nu}k^{\mu}k^{\nu}=\kappa T_{\mu\nu}k^{\mu}k^{\nu}$, and demanding that the NEC is always satisfied, $T_{\mu\nu}k^{\mu}k^{\nu}\geq 0$, 
we obtain the \emph{null convergence condition} (NCC):
 \be\label{eq2}
 R_{\mu\nu}k^{\mu}k^{\nu}\geq 0,\qquad
 \frac{d\theta}{d\lambda}+\frac{1}{2}\theta^{2}\le0
 \ee
 The above equations suggest that the converging null geodesics cannot start to diverge before meeting the 
{\it origin of coordinates}, or in other words converging null rays must meet the spacetime singularity 
in a finite time within GR, when the NEC is satisfied.

In a geometrically-flat spacetime, the obeyence of the convergence condition \eqref{eq2} results in the formation of a \emph{closed trapped surface} and subsequently, a singular spacetime. A closed trapped surface is formed when the ingoing and outgoing expansions of these null rays are negative, i.e. $\nabla_\mu k^\mu<0$.  However, in the case of a closed Universe in the $k=+1$ frame, the existence of closed trapped surfaces does not necessarily imply a singularity. See, \cite{Ellis:2003mb} for a full exposition. 
%In this regard, solving the singularity problem in geometrically-flat spacetime is indeed a hard problem.

Our aim in this paper is to use the power of the RE to show that {\it IDG}
can indeed yield the defocusing of null geodesic congruences, without violating the NEC, nor introducing ghosts or tachyons, in a geometrically-flat spacetime.

%%%%%%%%%%%%%%%%%%%%%%%%%%%%%%%%%%%%%%%%%%%%%%

\section{ Infinite derivative gravity}\label{IDG}
We shall now analyse how the RE is modified in the framework of quadratic curvature {\it IDG}. 
First, let us briefly recall the {\it IDG} action we are going to investigate, see~\cite{Biswas:2011ar}
\ba
\label{action}
S=\frac{1}{2}\int d^{4}x\sqrt{-g}\bigl(M_{P}^{2}R+
 R{\cal F}_{1}(\bar \Box)R+R^{\mu\nu}{\cal F}_{2}(\bar\Box)R_{\mu\nu}+ R^{\mu\nu\lambda\sigma}{\cal F}_{3}(\bar\Box)R_{\mu\nu\lambda\sigma}\bigr).
 \ea
Note that $\bar\Box = \Box/\Mc^2$ where $\Mc < M_P$ is the scale of modification of gravity in the UV. Current short distance tests of gravity, where 
there is no departure from Newtonian $1/r$ fall of gravity up to $5\times 10^{-6}$~m~\cite{Kapner}, place a rather mild limit on this scale, $\Mc\geq 10^{-2}$~eV~\cite{Edholm:2016hbt}. The non-local functions are defined by ${\cal F}_{i}({\bar \Box})\equiv \sum_{n=0}^{\infty}f_{i_n}(\Box/\Mc^2)^n$, where the coefficients $f_{i_n}$ are fixed by  the {\it ghost-free} condition, see ~\cite{Biswas:2005qr,Biswas:2013cha}~\footnote{Ref.~\cite{Biswas:2013kla} demonstrates how to recover various limits of action Eq.~(\ref{action}), at the level of graviton propagator,  
such as Weyl gravity, Gauss-Bonnet Gravity, Massive gravity, and scalar tensor theory. This paper also classifies which theory can be made
ghost free and which cannot.}.

In this paper, we will concentrate on linear perturbations around the Minkowski spacetime such that $$g_{\mu\nu}=\eta_{\mu\nu}+h_{\mu\nu},$$ where $\eta_{\mu\nu}$ is the Minkowski metric and $h_{\mu\nu}\equiv \delta g_{\mu\nu}$ is the metric tensor variation. 
%Upon application of the variational principle to the christoffel symbols and curvature, we can read off the relevant components of a gravitational theory up to linear order around Minkowski
%
%\ba
%R^{\rho(L)}_{\;\mu\sigma\nu} &=\frac{1}{2}\left(\partial_{\sigma}\partial_{\mu}h^\rho_{\nu}+\partial_{\nu}\partial^\rho h_{\mu\sigma}-\partial_{\nu}\partial_{\mu}h^\rho_{\sigma}-\partial_{\sigma}\partial^{\rho}h_{\mu\nu}\right)\,,\nn\\&
%\nn
%R^{(L)}_{\mu\nu}=\frac{1}{2}\left(\partial_{\sigma}\partial_{\mu}h_{\nu}^{\sigma}
%+\partial_{\nu}\partial_{\sigma}h_{\mu}^{\sigma}-\partial_{\nu}\partial_{\mu} h-\Box h_{\mu\nu}\right)\,,
%\\&
%\label{MinkR}
%R^{(L)}=\partial_{\mu}\partial_{\nu}h^{\mu\nu}-\square h\,,
%\ea
%
%where $L$ stands for linearised quantities and $\Box=\eta^{\mu\nu}\partial_\mu \partial_\nu$, in this case. 
The full equations of motion can be found in~\cite{Biswas:2013cha}.  At the linearised limit, 
the linearised equations of motion can be written in the following concise form
\[
\kappa T_{\mu\nu}	=	a(\bar\Box)R^{(L)}_{\mu\nu}-\frac{1}{2}\eta_{\mu\nu}c(\bar\Box)R^{(L)}-\frac{f(\bar\Box)}{2}\partial_{\mu}\partial_{\nu}R^{(L)}\,.
  \label{eomred}
\]
where $\Box=\eta^{\mu\nu}\partial_\mu \partial_\nu$,  $L$ stands for linearised quantities, and 
\ba
&a(\bar \Box)\equiv1+M_{P}^{-2}\left({\cal F}_{2}(\bar\Box)+2{\cal F}_{3}(\bar\Box)\right)\Box
 \nn
 \nn\\&
c(\bar\Box)\equiv1+M_{P}^{-2}\left(-4{\cal F}_{1}(\bar\Box)-{\cal F}_{2}(\bar\Box)+\frac{2}{3}{\cal F}_{3}(\bar\Box)\right)\Box
 \nn\\&
f(\bar\Box)\equiv M_{P}^{-2}\left(4{\cal F}_{1}(\bar\Box)+2{\cal F}_{2}(\bar\Box)+\frac{4}{3}{\cal F}_{3}(\bar\Box)\right),
   \label{abc}
 \ea
satisfying the following relations~\cite{Biswas:2011ar}
\begin{eqnarray}
-a(\bar\Box)+c(\bar\Box)+f(\bar\Box)\Box=0\,.
\end{eqnarray}
Taking the trace of the above equation  Eq.~(\ref{eomred}) yields:
\[\label{trace}
\kappa T=\frac{1}{2}\bigl(a(\bar\Box)-3c(\bar\Box)\bigr)R^{(L)}
.\]
We see here that the trace equation describes the scalar propagating mode, or precisely the spin-0 component of the graviton propagator~\footnote{For detailed computation of finding the spin projection operators and the graviton propagator for an {\it IDG}, see ~\cite{Biswas:2011ar,Biswas:2013kla}.}. This can be seen by studying the graviton propagator for the action Eq.\eqref{action}, by decomposing it in terms of true dynamical degrees of freedom around Minkowski spacetime:
\[
\label{MinkProp4}
\Pi(-k^{2})=\frac{\mathcal{P}^{2}}{k^{2}a(-k^{2})}+\frac{\mathcal{P}_{s}^{0}}{k^{2}\left(a(-k^{2})-3c(-k^{2})\right)}
,\] 
where we have suppressed the spacetime indices for clarity. To ensure freedom from ghosts,  both $a$ and $c$ must be {\it exponents of entire functions}. In particular, $a$ must not contain any roots in order to avoid the Weyl ghost~\cite{Biswas:2013kla}, while $c$ may contain at most one root, which would yield a scalar tensor theory of gravity. The addition of further poles to the function $c$ will necessarily be ghost-like or tachyonic, see \cite{Biswas:2005qr}, for further details. Moreover, the GR propagator must be recovered in the infrared limit as $k\rightarrow 0$, with $a(-k^2)\rightarrow1$ and $c(-k^2)\rightarrow 1$. For the choice $a=c$, we have
\begin{equation}
\label{MinkPropAC}
\Pi(-k^{2})=\frac{1}{a(-k^{2})}\biggl(\frac{\mathcal{P}^{2}}{k^2}-\frac{\mathcal{P}_{s}^{0}}{2k^2}\biggr).
 \end{equation}
which simply modulates the physical graviton propagator by an overall factor of $\sim 1/a(-k^2)$.

 \section{Defocusing conditions of Null Congruences for Infinite Derivative Gravity}\label{DNC}
 
Note that our action Eq.\eqref{action} will modify the RE, since $\kappa T_{\mu\nu}$ is modified, see Eq.\eqref{eomred}, and it is more insightful to first study the linearised limit. By contracting the energy momentum tensor 
with the congruences of null geodesics, $k^\mu$, we obtain
\[
\label{defocusgen}
R^{(L)}_{\mu\nu}k^{\mu}k^{\nu}=\frac{1}{a(\bar\Box)}\biggl[\kappa T_{\mu\nu}k^{\mu}k^{\nu}+\frac{1}{2}k^{\mu}k^{\nu}f(\bar\Box)\partial_{\mu}\partial_{\nu}R^{(L)}\biggr]
, \]
where the NCC requires $R^{(L)}_{\mu\nu}k^{\mu}k^{\nu}\geq0$.

Now let us study the simplest case, when the curvature is solely dependent on the cosmic time $t$, so that the D'Alembertian operator is given simply by $\Box=-\partial_t^2$.  The contribution of gravity to the RE is then:
 \[
 \label{defocus2}
 R^{(L)}_{\mu\nu}k^{\mu}k^{\nu}=\frac{1}{a(\bar\Box)}\biggl[\kappa T_{\mu\nu}k^{\mu}k^{\nu}-\frac{(k^{0})^{2}}{2}f(\bar\Box)\Box R^{(L)}\biggr]
. \]
To preserve the NEC, we must have $T_{\mu\nu}k^\mu k^\nu \geq 0$, whereas for the null rays to diverge, we require $R_{\mu\nu}k^\mu k^\nu<0$. Furthermore, from the propagator Eq.\eqref{propR2}, we require $a({\bar \Box})$ acting upon the curvature to be positive so as to avoid negative residues in the spin-2 component - a phenomenon known as the \emph{Weyl ghost}, i.e. $a({\bar \Box})R^{(L)}>0$.\\

\noindent
Taking all this into account, we obtain the \emph{minimal defocusing condition} for an {\it IDG} theory, which is independent of a background solution, and would yield a geodesically past-complete trajectory for a spatially-flat, homogeneous and isotropic background:
\[
\label{defocus3}
\frac{f(\bar\Box)\Box}{a(\bar\Box)}R^{(L)}> 0\Rightarrow \frac{a(\bar\Box)-c(\bar\Box)}{a(\bar\Box)}R^{(L)}> 0
,\]
with $T_{\mu\nu}k^\mu k^\nu=0$. From this defocusing condition, we may draw $3$ important conclusions:\\

\begin{enumerate}

\item{$a(\bar \Box)= c(\bar\Box)$: From Eq.\eqref{defocusgen}, defocusing of null geodesics can happen {\it only} when $a(\bar\Box)$ acting upon curvature is negative, which comes at the expense of the Weyl ghost~\cite{Biswas:2013kla}.}\\

\item{$a(\bar \Box)\neq c(\bar\Box)$: Since the curvature is always positive, we need a departure from the purely massless mode of the graviton propagator
in Eq.\eqref{MinkPropAC}. This condition tells us that in order for the null rays to defocus - a minimum requirement of a singularity-free theory of gravity - one requires an additional root in the spin-0 component of the graviton propagator. As $a(\bar\Box)$ does not introduce any new poles, the spin-2 component of the graviton
propagator remains massless. As such, one additional scalar degree of freedom must propagate in the spacetime besides the massless graviton, if we wish to satisfy the defocusing condition.}\\

\item{By inspection of Eq.~(\ref{abc}), it is possible to ``switch off'' any one of the form factors ${\cal F}_1,{\cal F}_2$ or ${\cal F}_3$ that make up the function $a({\bar \Box})$ and $c({\bar \Box})$, without altering the unitarity of the theory, see \cite{Biswas:2016etb,Biswas:2016egy}. The simplest choice is actually
${\cal F}_2=0$. In a conformally flat background such as in Minkowski space, de Sitter or indeed Freedman-Robertson-Walker metric, the Weyl tensor vanishes
in the background. Therefore, in this case, the action non-linear Eq.\eqref{action} reduces to the following form:
\ba
\label{action1}
S=\frac{1}{2}\int d^{4}x\sqrt{-g}\bigl[M_{P}^{2}R+ R{\cal F}_{1}(\bar \Box)R\bigr]\,.
 \ea
}
\end{enumerate}

Indeed, it is reassuring to note that from the most general covariant {\it IDG} action, Eq.\eqref{action}, we can deduce the simplest action of gravity
which has the potential to resolve the cosmological singularity problem, as given by Eq.~(\ref{action1}). Such a reduced action has been studied in~\cite{Biswas:2005qr}, where the authors found an {\it exact} non-singular bouncing solution for a homogeneous and istrotropic background for a  spatially-flat geometry. However, one must emphasise that the solution was indeed found based on a given {\it Ansatz}, which was invoked in order to solve the full non-linear equations of motion~\footnote{Such an {\it Ansatz}, taking the form $\Box R = c_1R+c_2$, where $c_{1,2}$ are constants and 
$R$ is the Ricci-scalar, was further verified in \cite{Barnaby:2007ve,Biswas:2006bs}. The sub-and super Hubble perturbations were also studied around the given bounce solution, and no instabilities were found~\cite{Biswas:2010zk}. }.
However, in the present work, our conclusion is deduced purely from geometric considerations of geodesics of null rays by analysing the RE.

Now, the relevant \emph{ghost-free condition} of the theory can be derived from the trace equation Eq.\eqref{trace}, where the functions $a(\bar\Box)$ and $c(\bar\Box)$ are recast in terms of 
another {\it exponent of an entire function} $\widetilde a(\bar\Box)$, containing no roots, by
\[
\label{GF1}
c(\bar\Box)=\frac{a(\bar\Box)}{3}\left[1+2\left(1-\alpha M_{P}^{-2}\Box\right){\widetilde a}(\bar\Box)\right]
.\]
Here, $\alpha$ is a constant, which a Taylor expansion of the trace equation Eq.\eqref{trace}, in conjunction with Eq.\eqref{abc}, reveals to be 
\begin{equation}
\alpha=6f_{1_0}+2f_{2_0}-\frac{M_P^2}{\Mc^2}\,. 
\end{equation}
The 
new graviton propagator can now be deduced by substituting Eq.\eqref{GF1} into Eq.\eqref{MinkProp4}, and subsequently decomposing the scalar propagating mode into partial fractions. 
\[
\label{Minkpropdec}
\Pi(-k^2)=\frac{1}{a(-k^2)}\biggl[\frac{{\cal P}^2}{k^2}-\frac{1}{2{\widetilde a}(-k^2)}\biggl(\frac{{\cal P}_s^0}{k^2}-\frac{{\cal P}_s^0}{k^2+m^2}\biggr)\biggr]
,\]
where we have defined the spin-0 particle with a mass, 
\begin{equation}\label{Staro-mass}
m^2\equiv \frac{M_P^2}{\alpha} > 0\,, 
\end{equation}
which must be positive to ensure that the mass is non-tachyonic, and non-zero to retain the essential new pole, as discussed previously. 
Armed with this, we are now in a position to describe the suitable defocusing condition which precludes the existence of ghosts. Substitution of Eq.\eqref{GF1} into Eq.\eqref{defocus3} leads to the linear condition;
\[
\label{defocus2}
(1-\Box/m^{2})\widetilde{a}({\bar \Box})R^{(L)}<R^{(L)}
   .\]
   
Before we conclude, we briefly extend our results to inhomogenous spacetimes. We may expand the general defocusing condition Eq.\eqref{defocusgen}, to include solutions, with spatial as well as temporal dependencies. Without loss of generality, we consider the perturbations along the $x$-direction, where $r=\sqrt{x^2+y^2+z^2}$.
As before, we require $T_{\mu\nu}k^\mu k^\nu \geq 0$ so as not to violate the NEC. We then read off the minimum requirement for the associated null rays to defocus
\[
\frac{f(\bar\Box)}{a(\bar\Box)}\left(\partial_{t}^{2}+\partial_r^2\right)R^{(L)}<0
,\]
where $\partial_r^2=\partial_i\partial^i$ is the Laplace operator. 
%The \emph{ghost-free} defocusing condition analgous to Eq.\eqref{defocus2}, is given by
%\[
%(1-\Box/m^{2})\tilde{a}(\Box)(\partial_{t}^{2}R^{(L)}+\partial_{r}^{2}R^{(L)})>\partial_{t}^{2}R^{(L)}+\partial_{r}^{2}R^{(L)}
%   .\]

%\section{Comparison with the Starobinsky Model}\label{Starobinsky}

As an additional note, a closer look at the above action Eq.\eqref{action1} suggests that if we were to consider $\Mc\rightarrow \infty$, the action would reduce to that of Starobinsky's model of 
inflation~\cite{Starobinsky}:
\[
\label{actionR2}
S_{R^2}=\frac{1}{2}\int d^4 x \bigl(M_P^2 R+f_{1_0}R^2\bigr).
\]
 Indeed, a curious question to ask is, could Starobinsky's action avoid the cosmological singularity?
At the limit $\Mc\rightarrow \infty$,  the propagator Eq.\eqref{Minkpropdec} can be expressed as
\[
\label{propR2}
\Pi_{R^{2}}=\Pi_{GR}+\frac{1}{2}\frac{\mathcal{P}_{s}^{0}}{k^{2}+m^{2}} ,
\]
where $m$ is given by Eq.~(\ref{Staro-mass}), with $\alpha=6f_{1_0}\geq 0$, and to avoid tachyonic mass, $m^2>0$.
 However, the fundamental difference can be seen by comparing the propagator for $R^2$-gravity with an IDG propagator, see Eq.\eqref{Minkpropdec}.  In the local limit, $a={\widetilde a}\rightarrow 1$. Furthermore, as we are making comparisons with the propagator in momentum space, the D'Alembertian takes the form $\Box\rightarrow -k^2$. 
Taking these limits on the defocusing condition Eq. \eqref{defocus2} reveals the analogous condition for the Starobinsky model
\[
(k^2/m^2) R^{(L)}<0
.\]
In this scenario, in order to avoid convergence, we have two options: either 1.) the spin-0 particle has imaginary mass, i.e. $m^2<0$, or 2.) the curvature is negative.
\begin{enumerate}
\item A particle/field with imaginary mass is, by definition, tachyonic. As $m^2=M_P^2/6f_{1_0}$, where $f_{1_0}$ is the coefficient attached to the $R^2$ term in the action Eq.~\eqref{actionR2}, this corresponds to the $S\sim R-R^2$ form of the Starobinsky model. 
\item If one dictates that this coefficient is necessarily positive, thus avoiding tachyons, the defocusing condition may be satisfied if one allows for negative curvature.  However, negative curvature would contradict the requirement of accelerated expansion of the Universe, which is vital to realise primordial inflation. This corresponds to the $S\sim R+R^2$ form of the Starobinsky model.
\end{enumerate}
As such, the Starobinsky model cannot pair inflation with resolving the Big Bang Singularity.

\section{Conclusion}

From very generic considerations of a covariant, higher derivative theory of gravity, one can conclude that 
null congruences can be made complete, or can be defocused, provided two essential criteria are satisfied 
at the microscopic level, i.e. in terms of the graviton propagator: 1) 
The graviton propagator must contain a scalar propagating mode which contains one additional root,
besides the massless spin-2, i.e. two helicity states; and 2) the infinite derivative theory of gravity must be, at least, ghost and tachyon free. 
This conclusion verifies earlier analyses, which were dependent on an {\it Ansatz}-led, time-dependent solution to the equations of motion for an {\it IDG} theory, which described a non-singular cosmology in an homogeneous, isotropic and geometrically-flat framework.

This entails that an \emph{infinite} covariant derivative  theory of gravity which can avoid ghosts or tachyons in the graviton propagator,
can yield a non-singular bouncing cosmology,  and a possible UV completion of  original Starobinsky inflation~\cite{Biswas:2013dry,Craps:2014wga,Koshelev:2016xqb}. In future, we shall be able to generalise our current discussion to de Sitter  backgrounds, based on our recent
analysis of propagator in de Sitter spacetimes~\cite{Biswas:2016etb,Biswas:2016egy}. We can also attempt to analyse the defocusing of null congruences for a generic inhomogeneous and anisotropic backgrounds, however, one requires to first analyse the true dynamical propagating degrees of 
freedom in such backgrounds, and the graviton propagator.  Finding the latter would be a subject of discussion by itself, and we leave these discussions for future publication.

\section*{Acknowledgments}
We would like to thank Tirthabir Biswas, Valery Frolov, Terry Tomboulis, Alexei Starobinsky for helpful discussions. AC is funded 
by STFC grant no ST/K50208X/1. AK is supported by the FCT Portugal fellowship SFRH/BPD/105212/2014 and in part by RFBR grant 14-01-00707. The work of A.M. is supported in part by the Lancaster-Manchester-Sheffield Consortium for Fundamental Physics under STFC grant ST/L000520/1.
%%%%%%%%%%%%%%%%%%%%%%%%%%%%%%%%%%%%%%%%%%%%%
%%%%%%%%%%%%%%%%%%%%%%%%%%%%%%%%%%%%%%%%%%%%%%

\end{document}